\documentclass[runningheads]{llncs}
\usepackage[T1]{fontenc}
\usepackage{graphicx}
\usepackage{adjustbox}
\usepackage{amsmath,amssymb,amsfonts}
\usepackage{algorithm,algorithmic}
\usepackage{graphicx}
\usepackage{textcomp}                         
\usepackage{hyperref}
\usepackage{soul}
\usepackage{color}
\usepackage{multirow}
\usepackage{booktabs}
\usepackage{threeparttable}
\usepackage{graphicx}

\begin{document}
\title{A Game Theoretical vulnerability analysis of Adversarial Attack}

\titlerunning{Game Theoretical analysis of Adversarial Attack}

\author{Khondker Fariha Hossain \inst{1} \and 
  Alireza Tavakkoli \inst{2} \and Shamik Sengupta \inst{3}}
% %
\authorrunning{K.F. Hossain, et al.}
\institute{University of Nevada, Reno, NV 89557 \\ \email{khondkerfarihah@nevada.unr.edu \inst{1}}, \email{tavakkol@unr.edu \inst{2}},
\email{ssengupta@unr.edu \inst{3}}}


%
\maketitle              % typeset the header of the contribution
\begin{abstract}
In recent times deep learning has been widely used for automating various security tasks in Cyber Domains. However, adversaries manipulate data in many situations and diminish the deployed deep learning model's accuracy. One notable example is fooling CAPTCHA data to access the CAPTCHA-based Classifier leading to the critical system being vulnerable to cybersecurity attacks. To alleviate this, we propose a computational framework of game theory to analyze the CAPTCHA-based Classifier's vulnerability, strategy, and outcomes by forming a simultaneous two-player game. We apply the Fast Gradient Symbol Method (FGSM) and One Pixel Attack on CAPTCHA Data to imitate real-life scenarios of possible cyber-attack. Subsequently, to interpret this scenario from a Game theoretical perspective, we represent the interaction in the Stackelberg Game in Kuhn tree to study players' possible behaviors and actions by applying our Classifier's actual predicted values. Thus, we interpret potential attacks in deep learning applications while representing viable defense strategies in the game theory prospect.

\end{abstract}

\section{Introduction}

In machine learning, a classification model refers to a predictive modeling problem, assuming that the training and testing data is generated from the same underlying distribution. But in real life, data evolves, and sometimes malicious instances\cite{szegedy2013intriguing} change the drift of the existing data. These perturbations in the input data make machine learning models vulnerable and result in unusual changes in output data. Though this modification does not affect the human perceptual ability to identify them, deep learning models, on the other hand, are quite susceptible, creating a threat to security and safety concerns.

Researchers have suggested many additive defense methods in response to such additive attacks. The current defenses against such attacks are training modified data focusing on gradient pathways\cite{kurakin2016adversarial}, performing different filtering, or removing adversarial perturbation from the input data\cite{meng2017magnet}. Though the variation of the adversarial attacks (both targeted and non-targeted) changes the metric of the classifier, it can be identified by figuring out the perturbation by calculating the gradient.
Each time a new attack strategy has been proposed, the defense strategy for the such attack came within a small margin of time. Moreover, novel defenses are being introduced to improve upon existing techniques to fend off each new attack. This back-and-forth game between attack and defense is persistently recurring, indicating reaching a consent resolution. As the attack's synopsis and the defense are robust, it became significant to understand the pattern and behavior of the attacker to create a robust defense. Many of the defense techniques incorporate popular decision-making-based frameworks to explore the interactions between the attacker and the defender. One such popular tactic is applying a mathematically based model like Game theory\cite{fudenberg1991game}. Game theory is a mathematical framework that analyzes the behavior of the players of the game. In the cyber domain, the players are the attacker and the defender. The tradeoff between the attacker and the defender is the cost of adopting the strategies. In deep learning models, the classifiers are considered the defender, while adversarial attacks play the attacker role. So the computational tradeoff between the adversarial attack and the classifier is distributed based on the gain from the attack.

One popular method, the Game theory, has been extensively applied in cybersecurity \cite{zhang2009impulse} vulnerabilities to capture the interaction between players quantitatively. Game theory is the study of mathematical models for analyzing the decision-making process for rational agents \cite{myerson2013game}. Game theory is successfully applied in the security-based scenario as one of the core qualities of the Game theory perspective is to analyze the cognitive behavior of the players\cite{camerer2011behavioral}. Before executing a Machine Learning model for Cloud infrastructure, it is required to analyze not only the implementation or computation cost but also the recovery or vulnerability cost. In this synopsis, the Game theory is a proper solution. Again, though we consider players to be rational in a Game Theoretic sense, in real life, players can be irrational. Moreover, sometimes while designing a Game, many auxiliary players are not included in the strategy that might affect the game's outcome. Reinforcement learning is a type of Game that helps distinguish different players in real life and is beneficial in real-life scenarios \cite{nowe2012game}.

In this paper, our main contribution can be summarized below:

\begin{itemize}
  \item We empirically show that Deep Neural Networks (DNN) used in cyber domains are vulnerable to adversarial attacks. 
  \item To eliminate this enigma, it is vital to do a behavioral analysis of the attacker's and defender's perspectives to come out with a more subtle way of handling these attacks. 
\end{itemize}

\begin{figure*}[!t]
    \centering
    \includegraphics[width=0.6\linewidth]{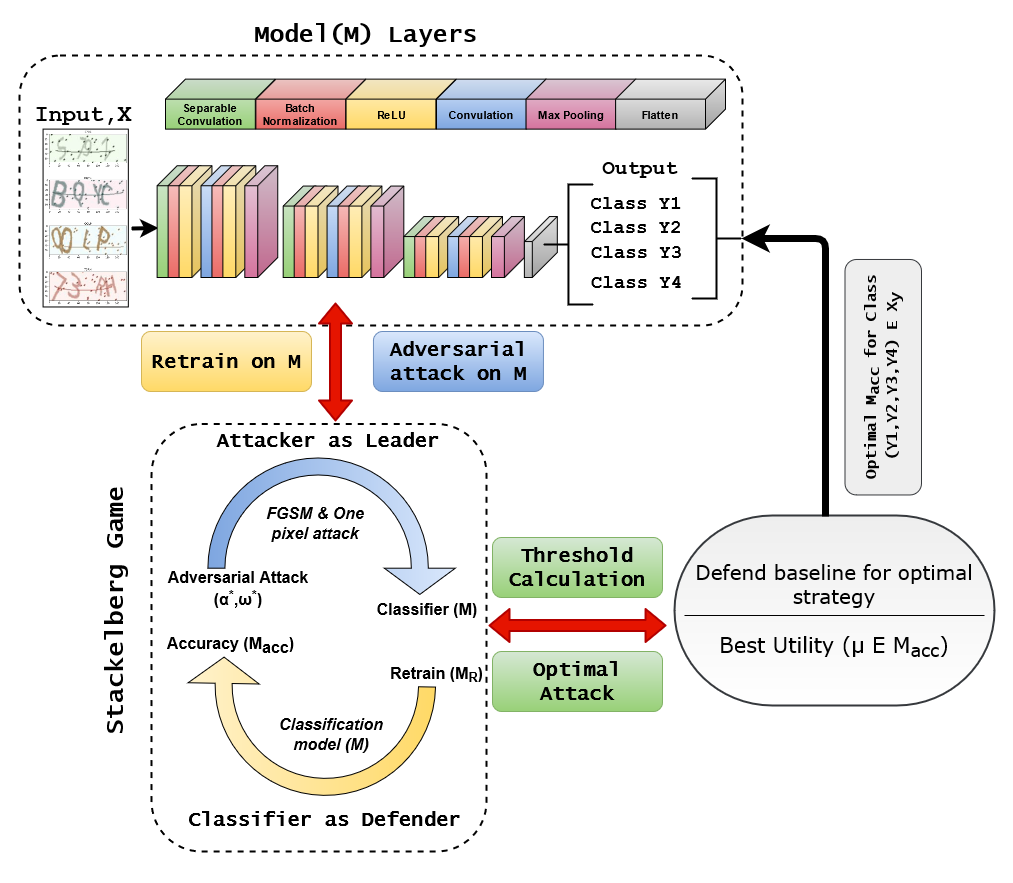}
    \caption{ Representation of the Stackelberg game for the CNN model which is consist of three combined layers to decode in three steps.The generated CAPTCHA images are the input while the prediction outputs are the four classes.}
    \label{fig1}
\end{figure*}

\section{Methodology}
\label{met}

\subsection{CNN model and Loss Function }
\label{subsec:cnnmodel}

Fig. 1 illustrates the Deep Convolutional Neural Network(CNN) architecture we propose to classify CAPTCHA images. 
The CNN model takes the CAPTCHA images x as input and  predicts the classes of the original input. The model consists of multiple Convolution, Batch-Norm, ReLU, and Fully Connected layers. We use convolution and separable convolution for downsampling three times. For convolution we use kernel size, k = 3, stride, s = 1 and padding, p = 0, except for downsampling convolution where we use stride, s = 2. The convolution layer is succeeded by Batch-normalization, ReLU, and Max-pooling layers. After that, we use one fully connected layer. The encoder consists of 9 convolution layers and four dense layers. The number of features are for each layer is [E1;E2;E3;E4;E5;E6;E7;E8;E9;E10]= [32; 32;64;64; 128; 128; 256; 64; 32]. We use one output activation: classification with Softmax for four class prediction. For the classification of the particular characters of the CAPTCHA image, we use categorical cross-entropy \cite{zhang2009impulse} as in Eq.~\ref{eq2}.
\begin{equation}
    \mathcal{L}_{class}(D) = -\sum^{k}_{i=0} y_i\log(y'_i)
\label{eq1}
\end{equation}

\subsection{FGSM Attack}
\label{subsec:fgsmattack}

\begin{figure}[htp]
    \centering
    \includegraphics[width=0.8\linewidth]{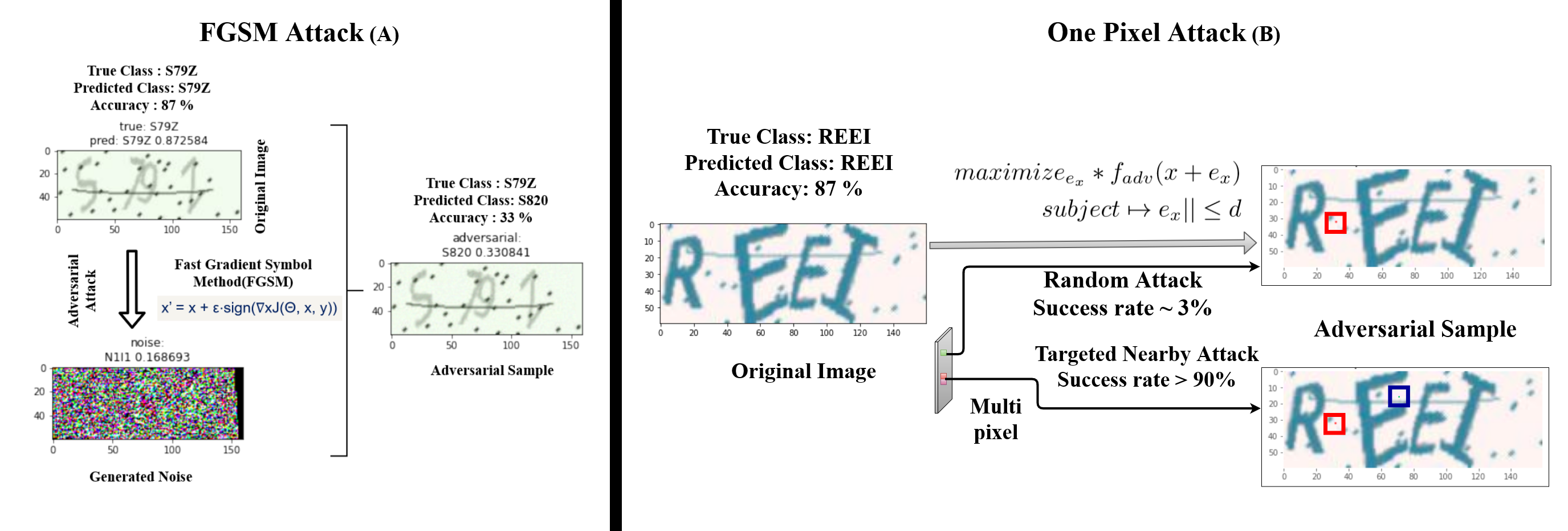}
    \caption{(A)FGSM attack on the original image and the generated adversarial sample is predicted with low accuracy, (B)One pixel attack on the original image with the changes on a single pixel and multi pixels which generate the adversarial sample with low prediction accuracy}\label{fig2}
\end{figure}

To create an adversarial sample, the Fast Gradient Sign Method (FGSM)\cite{goodfellow2014explaining} uses the gradients of the neural network. The attack mechanism is to calculate the gradients of the loss based on input images to create a new image that maximizes the loss\cite{madry2017towards}. We illustate the FGSM attack in Fig.~\ref{fig2}.
\iffalse
The distinct part of this attack is the consideration of the gradient in respect of the original image to maximize the loss. So the procedure implies calculating the contribution of each image pixels towards the loss function by using the chain rule and calculate the corresponding gradient. For the model that is not trained, the gradient is not taken into consideration; thus, it will be constant. Therefore, The FGSM attack only targets the trained model to fool.
\fi
The FGSM attack is given in Eq.~\ref{eq2}.

\begin{equation}
   x'=x+\epsilon \cdot sign(\bigtriangledown xJ(\theta,x,y ))
    \label{eq2}
\end{equation}

In Eq.~\ref{eq2} $x'$ symbolized the optimal disturbance or the adversarial sample. Here, $x$ represents the Original input image, and $y$ represents the corresponding label. $\theta$ is the model parameter, so the loss function, $J$, is calculated based on the value of $\theta$, $x$, and $y$. Again the perturbation level $\epsilon$ is multiplied to ensure a small perturbation. 

\subsection{ One Pixel Attack }
\label{subsec:onePixelAttack}
A single-pixel modification can make extremely deep Neural Networks like Resnet, VGG, and other such models vulnerable \cite{su2019one}. Moreover, in the paper, we manifested that initial local perturbation shows the characteristics of spreading and becoming global, which emphasized that nearby pixels of the targeted perturbed pixel are exposed to the same vulnerability. We illustrate the one pixel attack in Fig.~\ref{fig2}.

\iffalse
\begin{figure}[!b]
    \centering
    \includegraphics[width=1\linewidth]{Fig3.png}
    \caption{One pixel attack on the original image with the changes on a single pixel and multi pixels which generate the adversarial sample with low prediction accuracy}\label{fig3}
\end{figure}
\fi
To understand the mathematical intuition behind the one pixel attack, a function $f(x)$ that takes images as input where $f(x)$ $\epsilon$ $R$ as in $R$ is the distribution of the input image. The output of the classifier is the probability of the label, $R^l$ where $x$ $\epsilon$ $R^ {p*q}$. Here, $R$ stands for real numbers, $l$ is the labels of the image and $p$,$q$ are the dimension of the input image. To create an adversarial sample $x'$ we will put some perturbation $\varepsilon_{x} \epsilon R$ in the input image $x$. 
\iffalse
\begin{equation}
   x' = x + \varepsilon _{x}'
    \label{eq3}
\end{equation}
\fi
\begin{equation}
   {x'\epsilon R^{n} | argmax (f(x'))\neq argmax(f(x))}
    \label{eq4}
\end{equation}

\begin{equation}
\begin{split}
   maximize_{\varepsilon _{x}} f_{adv}(x+\varepsilon _{x})_{c}\\
   subject \mapsto ||\varepsilon _{x} 
    \label{eq5}
\end{split}    
\end{equation}
\iffalse
In the case of untargeted attack, the soft-label $f_{adv}(x)_{c}$ need to be minimized.
\begin{equation}
\begin{split}
   maximize_{\varepsilon _{x}} - f_{adv}(x+\varepsilon _{x})_{c}\\
   subject \mapsto ||\varepsilon _{x}
    \label{eq6}
\end{split}    
\end{equation}
\fi
So the equation for the one pixel attack can be represented as in Eq.~\ref{eq7}. $d$ is the small value that modify the number of pixels are going to attack. In case of one pixel attack $d=1$ which will only manipulate a single pixel.  

\begin{equation}
\begin{split}
   maximize_{e_{x}}* f_{adv}(x+e_{x})\\
   subject\mapsto e_{x} || \leq d  
    \label{eq7}
\end{split}    
\end{equation}

\subsection{Optimization: Differential Evolution(DE)}
\label{subsec:DE}

Differential evolution is a population-based heuristic optimization algorithm that is used here to encode the perturbation optimally. 
Furthermore, for the one-pixel attack, the mechanism of creating perturbation with five elements, x-y coordinate, and RGB value makes it independent of the classifier type, and by applying the DE for the attack, it is sufficient to know the probability labels. q. Eq.~\ref{eq8}\cite{su2019one} is the formulation of the DE, where $x_{i}$ is an element of the candidate solution, $r_{1},r_{2},r_{3}$ are random numbers,$F$ is the scale parameter set to be $0.5$, $g$ is the current index of generation.

\begin{equation}
x_{i}(g+1) = x_{r1}(g)+F(x_{r2}(g)-x_{r3}(g)), r1\neq r2\neq r3
    \label{eq8}
\end{equation}

\section{Experimental Analysis}
\label{expe}
\subsection{Dataset}
\label{subsec:dataset}
In our experiment, we used CAPTCHA dataset to analyze the attack. CAPTCHA is a type of challenge-response test applied in computing to determine if the user is human or not\cite{wang2017recognition}. In the experiment, we generated 20000 images where we allowed to have four characters in the image. The characters will be chosen in-between (A-Z), all in uppercase, and the numbers from (0-9). All the characters are randomly placed in the image. The library optimally avoids generating the same images, and in our dataset we made sure not to have any duplicate CAPTCHA image. In \ref{fig4}, you can see the instance of three generated CAPTCHA images.

\subsection{Hyper-parameter tuning}
%\vspace{-3mm}\subsection{Hyper-parameter tuning}\vspace{-1mm}
\label{subsec:hyper}
For training, we used categorical cross entropy loss \cite{zhang2009impulse} . For optimizer, we used Adam \cite{kingma2014adam}, with learning rate $\alpha = 1e-4$. We train with mini-batches with batch size, $b=32$ for 20 epochs. It took approximately 1.5 hours to train our model on NVIDIA P100 GPU.

\subsection{Training procedure}
\label{subsec:training}
Firstly, we split the dataset into train and test, where the test portion contains 20 percent of the data. Then using the provided Hyperparameters, we trained on the training dataset. We used Accuracy matrices to predict the test data. Later we performed the FGSM Attack and One-pixel attack on the same test dataset and calculated the accuracy. If the accuracy was greater than 50\%, we retrained the classifier on the same test data, but this time we included the adversary while training. In the next step, we calculated the accuracy, and therefore, if it is less than 85\%, we did the retrain again. The condition of the retrain the model is, it has to be less than the threshold, and after the adversarial attack, the classifier must include all the generated adversarial samples, including the original input. 

\begin{algorithm}[!t]
\caption{Adversarial Training based Stackelberg Game}
\label{alg1}
\begin{algorithmic}[1]
 \renewcommand{\algorithmicrequire}{\textbf{Input:}}
 \renewcommand{\algorithmicensure}{\textbf{Output:}}
  \REQUIRE $x^{i} \epsilon X$, $y^i \epsilon Y$;Labelled training data $X_{train}$, Labelled testing data $X_{test}$
 \ENSURE $M(Accuracy)$
  \STATE \textbf{Initialize hyper-parameters}: \\$max\_epoch$, $max\_batch$, $\omega_{M}$, $\alpha_{M}$,  $\beta^1_{M}$, $\beta^2_{M}$
  \STATE Train Model($M$) on $X_{train}$ to get output model accuracy $M_{acc}$\textit{(default)}
  \STATE Adversarial attacks $F\mapsto FGSM$ or $O\mapsto One-pixel-attack$ with perturbation $(\alpha)$
  \FOR{$e=1\ to\ max\_epoch$}
  \FOR{$b=1\ to\ max\_batch$}
    \STATE $\mathcal{L}_{class}(M)= -log(\frac{e^{y_{p}}}{\sum_{i}^{k}e^{y_{k}}})|| y_{p}\subseteq(+y_{class})$ 
    \STATE $\mathcal{L}_{class}(M) \gets M(x),y$
    \STATE $\omega_{M} \gets \omega_{M}+ Adam(M,\omega_{M},\alpha_{M},\beta^1_{M},\beta^2_{M})$
    \STATE Save weights($\omega$) and snapshot of $M$
    \ENDFOR
  \ENDFOR
\STATE Evaluate $M_{acc}$ for $X_{test}$ of $M$  
\IF{$M_{acc}$ $>$ $50 \%$} 
    \STATE Retrain on $M$ with Adversarial sample $(\alpha^*,\omega^*)\epsilon{L}_{class}(M)$
    \STATE $\mathcal{L}_{class_{1}}(M)= -log(\frac{e^{y_{p}}}{\sum_{i}^{k}e^{y_{k}}})|| y_{p}\subseteq(+y_{class(\alpha^*,\omega^*)})$ 
    \STATE $\mathcal{L}_{class_{1}}(M) \gets M(x),y$
    \STATE Calculate $M_{acc}$ for $X_{test}$ $\cup $ $X_{test}+X_{test\alpha^*}$  of $M$
    \ENDIF
\STATE Evaluate $M_{acc}$ for $X_{test}$ of $M$ (2nd step)  
\IF{$M_{acc}$ $<$ $85 \%$} 
    \STATE Retrain on $M$ with Adversarial sample $(\alpha^*,\omega^*)\epsilon[{L}_{class}(M)+{L}_{class_{1}}(M)]$
    \STATE $\mathcal{L}_{class_{2}}(M)= -log(\frac{e^{y_{p}}}{\sum_{i}^{k}e^{y_{k}}})|| y_{p}\subseteq(+y_{class_{2}}(\alpha^*,\omega^*))$ 
    \STATE $\mathcal{L}_{class_{2}}(M) \gets M(x),y$
    \STATE Calculate $M_{acc}$ for $X_{test}$ $\cup $ $X_{test}+X_{test\alpha^*}$  of $M$
    \ENDIF
\STATE return $M_{acc}$    
\end{algorithmic}
\end{algorithm}

\begin{figure*}[htb]
    \centering
    \includegraphics[width=0.6\linewidth]{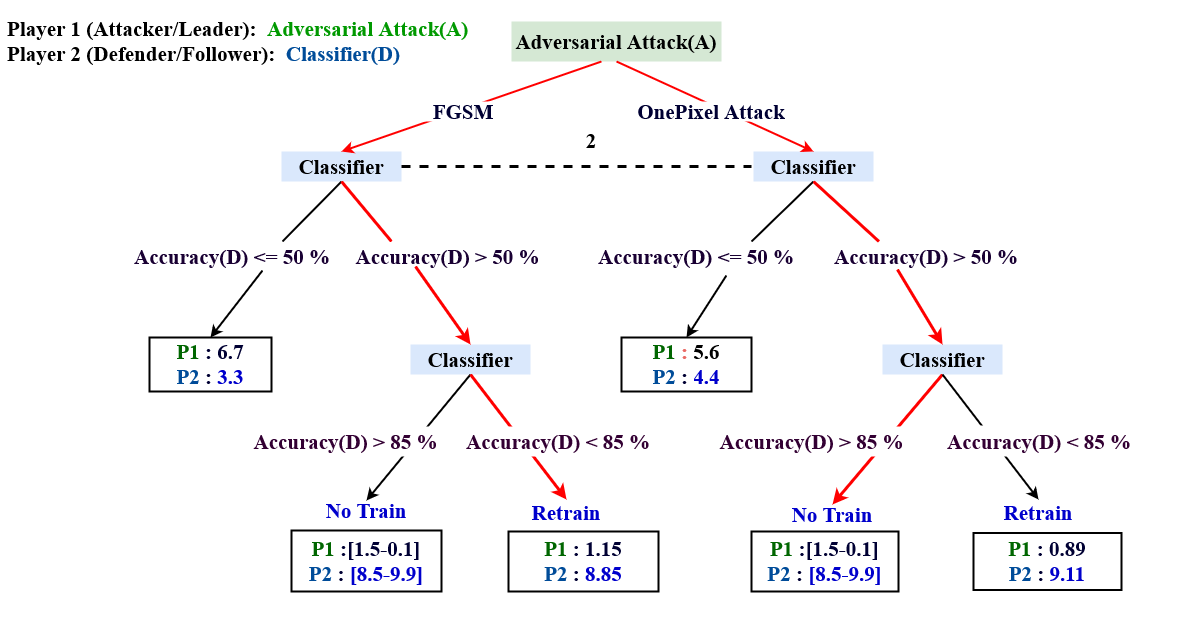}
    \caption{The representation of the Game's kuhn tree between the Attacker (Adversarial Attack) and the Defender (Classifier) to calculate the Optimal solution }\label{fig4}
\end{figure*}

\section{Analysis: Game Theory}
\label{game}
%\vspace{-3mm}\section{Conclusion}\vspace{-1mm}
 
 The game is designed with four possible outcomes. These outcomes are labeled with payoff from 0 through 10 for both of the players. The outcome value is 0 if only there is no attack from the player's perspective. The payoff and strategies for both players are given below: 

The game is designed with four possible outcomes. These outcomes are labeled with payoffs from 0 through 10 for both players. The outcome value is 0 if there is no attack from the player's perspective. The payoff and strategies for both players are given below: 

 \begin{itemize}
   \item Initially, we assume that, if there is no attack, the total payoff of the game is 0.
   \item The Attacker can choose any of the adversarial attacks, FGSM or One-pixel attack, but the classifier will initially classify both cases
   \item  For a given two class labels, it is a binary classification problem for the classifier. In \ref{fig4}, the Kuhn tree is designed as a binary tree; as per probability distribution, a discrete output of the binary classifier is $X \epsilon (0,1)$. Therefore based on the accuracy of the classifier, it can be branched for more or less than $50\%$.
   \item If the classifier's output is less than $50\%$, the Attacker is already successful and will get a payoff of more than $50\%$. In the \ref{fig4} we can see that for both FGSM and One-pixel attacks, the Attacker got more value, and the defender will get the rest payoff as \textit{(1- Attacker's payoff)} which is less than Attacker's payoff.
   \item For the second scenario, where the accuracy is greater than $50\%$, the classifier will defend by classifying with higher certainty. If the accuracy is now more than $85\%$, then the classifier will not have to retrain on the adversarial sample as it is familiar with the adversarial data distribution. However, suppose the classifier predicts with less than 85\% certainty. In that case, It will utilize this data as an adversarial sample and train again on the original data along with the adversarial sample.
   \item Finally, the classifier's target is to get an accuracy of more than $85\%$. If the classifier, as in defender, brings more value, the game ends or follows the previous step. The core purpose of the game is to win the classifier (Defender). As in the generalized version in the case of the DNN-based classification task for the cyber domain, we can consider that achieving an accuracy of more than $90\%$ is reliable.
  
 \end{itemize}

\subsection{Game Formulation:}
\label{subsec:player}
We formulate the game between the attacker and classifier as a two-player Stackelberg game \cite{zhang2009impulse}. The fundamental components of forming the game are: modeling the actions taken by the players, formatting players as the decision-makers, calculating the payoffs for each players. While forming the Stackelberg game, we consider the adversarial attacker's action as the Leader (L) who initiates the game by manipulating data over a strategy space.  In response to each attack, the classifier is considered as a Follower (F) who manipulates training data distribution based on the re-optimization of the weights.

\begin{figure*}[htb]
    \centering
    \includegraphics[width=0.9\linewidth]{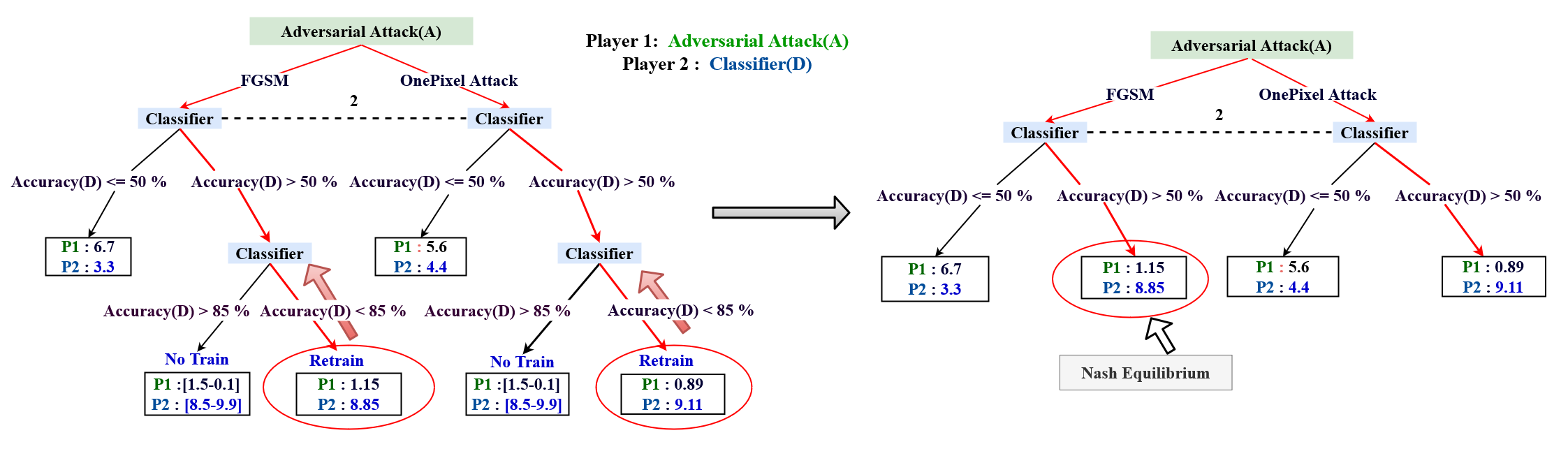}
    \caption{The Optimal solution is calculated from the Game's kuhn tree between the Attacker (Adversarial Attack) and the Defender (Classifier)}\label{fig5}
\end{figure*}

As per the rule of the Stackelberg game, the leader(L) will initiate the game by the adversarial attack. Here, there are two action spaces, two attacks $F \mapsto FGSM$ and $O \mapsto One-pixel-attack$, and feature space of the weight,$W$, would initially be chosen randomly. L's and F's payoff function will settle through $J_{L} \epsilon R$ and $J_{F} \epsilon R$. After the attack, the classifier will try to learn the optimal parameter $w^*\epsilon W$ from the training set based on the adversary $ \alpha^* \epsilon [F,O]$. The optimal adversary will estimate  through the payoffs $J_{L}$ and $J_{F}$ by solving Eq.~\ref{eq9}

\begin{equation}
\begin{gathered}
   \alpha = \arg\max_{\alpha \epsilon [F,O]}J_{L}(\alpha,\omega ^{*}) \\
   \omega = \arg\max_{\omega \epsilon W}J_{F}(\alpha^{*},\omega) 
    \label{eq9}
\end{gathered}
\end{equation}

The game is Stackelberg simultaneous game, so we formulate the game from Eq.~\ref{eq9} to Eq.~\ref{eq10}. Therefore the attacker(L) starts the game by the adversarial attacks then the classifier receives the new data space along with the original data $x\epsilon X$, including the actions and weights$(\alpha^*,\omega^*)$.

\begin{equation}
    (\alpha^{*},\omega ^{*}) = \arg\max_{\alpha \epsilon [F,O]}J_{L}(\alpha,\arg\max_{\omega \epsilon W}J_{F}(\alpha,\omega))
    \label{eq10}
\end{equation}

The game is a two-class classification problem with diverging data distributions, and for theoretical perspective, we considered the data space as a one-dimensional feature space. We denote the distribution $\sigma$ to determine the payoff relation between L's and F's by calculating the loss function, Eq.~\ref{eq11}. In our model, we used categorical cross-entropy as a loss function, and we calculate the cost, $\mathbb{C}_{F}$  for follower and the cost, $\mathbb C_{L} =(1- \mathbb C_{F})$ for the leader to determine the payoff $J_{L}$ and $J_{F}$.

\begin{equation}
\begin{split}
    \mathbb{C}_{F} = -\sum_{i=0}^{k}x_{i} log (x'_{i}) \\
    J_{L}+J_{F} = \sigma + \mathbb{C}_{L}(\alpha,\omega)+\mathbb{C}_{F}
    \label{eq11}
\end{split}
\end{equation}

From Eq.~\ref{eq9} and Eq.~\ref{eq10} for the two-player Stackelberg game, we can express the objective function Eq.~\ref{eq11} for simultaneous actions. In Fig.~\ref{fig4}, we can see that the type of attack is unknown to the classifier in the initial stage, so there will not be any changes in the weights and parameters; thus, its an incomplete information game. In the following stages, depending on the attack type, the classifier will get updated weights ($\omega$) and parameters($\alpha$) along with relative payoff value $\mathbb C_{F}$ and $\mathbb C_{L}$ for $J_{L}$ and $J_{F}$. The final objective function is given in Eq.~\ref{eq12}.

\begin{equation}
(\alpha^{*},\omega ^{*}) = \arg\max_{\alpha \epsilon F,O}J_{L}(\alpha,\arg\max_{\omega \epsilon W}(\sigma + \mathbb{C}_{L}(\alpha,\omega)+\mathbb{C}_{F}-J_{F}(\alpha,\omega)))
    \label{eq12}
\end{equation}

\subsection{A Characterization of Threshold for Game Strategy}
\label{subsec:Threshold}
 We have rational agents playing the game, all the win or loss scenarios for the attacker were illustrated based on manipulation of Retrain defense strategy. To demonstrate this, we first established that the Threshold must be categorized as a Binomial probability distribution in the initial attack. Then we set that the Threshold value of the defender should have more than 85\% accuracy against the first attack for a successful defense. 

\begin{itemize}
    
\item \textbf{$M_{acc}$ as Threshold}
For all the input images $x \epsilon X$, the Accuracy rate ($M_{acc} \epsilon \mathbb{R}$)  will act as Threshold for all possible scenario, to measure the success rate of the attacker and the defender in the classification model. 

Following Probability distribution addressed the first attacks on the classification model, and for $M_{acc}$ as Threshold, it outlines the Threshold value based on the probability distribution.   

\vspace{1mm}
\item \textbf{Binomial probability distribution for the First attack}
For any $x \epsilon X $ the success of the attacker is $f(s)$.

\begin{equation}
    \begin{split}
        f(s|A,p) =\begin{pmatrix} A\\ s\end{pmatrix}p^{s}(1-p)^{A-s}\\
        f(s) =\left\{\begin{matrix}
Successful_{A} \mapsto f(s|A,p) \leq 50\% \\ Not-Successful_A \mapsto f(s|A,p)> 50\% \end{matrix}\right.
    \label{eq13}
    \end{split}
\end{equation}

The numbers $A$ and $p$ signifies the distributions: the number of attacks and the probability of success in each Attacks. So the probability of s number of success in $A$ number of attacks is represented by the function $f(s|A,p)$ in Eq.~\ref{eq13}. The Binomial Coefficient $(As)$ measures the possible success $s$ can be obtained from $A$ attacks without replacement wheres $p^{s}(1-p)^{A-s}$ estimates the probability of such outcomes.

For the initial attack, the Binomial distribution is acceptable as it only deals with the success of the attack. So the probable chance for the function is, $f(s|A,p) \mapsto (<50\%>) $: either it will be a successful attack or unsuccessful, nothing other then that. In our scenario, for the attack to be successful, $f(s|A,p)> 50\%$,  the classifier's weights, and the parameters will have to change as well as the payoff $J_{L}$ and $J_{F}$, and can be validated from from Eq.~\ref{eq11} and Eq.~\ref{eq12}. For the further attacks on the next stages, it will follow the following Retraining as a successful Defense.

\vspace{1mm}
\item \textbf{Retrain as a successful Defense for higher $M_{acc}$}
For any attacks after the first attack, the success of $f_{D}(s)$ will be greater or less than the original $M_{acc}$ and Retrain$(\alpha^{*},\omega^{*})$ as the defense will be continued until $f(s^{(\alpha^{*},{\omega^{*}})}_{M_{acc}})> M_{acc}$.

\begin{equation}
    \begin{split}
f(s^{(\alpha^{*},{\omega^{*}})}_{M_{acc}}) =\left\{\begin{matrix}
Successful_D \mapsto {M_{acc}}^{(\alpha^{*},\omega^{*})} > 85\% \\ Not-Successful_D \mapsto {M_{acc}}^{(\alpha^{*},\omega^{*})} < 85\% \end{matrix}\right.
    \label{eq14}
    \end{split}
\end{equation}

From the following attacks, the objective is to Retrain model$(M)$ on the Adversarial samples $X^{(\alpha^{*},\omega^{*})}\epsilon X$ to gain the original accuracy $M_{acc}^{Original}\epsilon \mathbb{R}$. In our experiment, we retained on updated sample $X^{(\alpha^{*},\omega^{*})}$ along with the original sample ${X}$ to gain the original $M_{acc}^{Original}> 85\%$  

\end{itemize}

\subsection{Utility Value and Simulation Results on CAPTCHA Data}
\label{subsec:utility}

To evaluate the Deep Learning Model on the classification of the CAPTCHA dataset, We first trained the model on $X_{train} \epsilon X$ and calculated $M_{acc}^{Original}$ on the test data, $X_{test}$ for the \textit{Threshold} measurements. For computational tractability, the first attack in the Stackelberg Game is a Black Box attack. So initially, it is an incomplete information game; therefore, the attack type and the number of attacks are unknown. In Fig~\ref{fig4}, the \textit{Kuhn tree}, we can see the Classifier is unknown to the attacks \textit{FGSM} and \textit{One-Pixel-Attack} along with the parameters($\alpha$) and weights($\omega$) changes after the attack.

\begin{table}[!h]
\centering
\caption{\textbf{Utility as $M_{acc}$} : The accuracy value of the Model is represented as Utility in two stages for both attacks.}
\begin{adjustbox}{width=0.6\linewidth}
\begin{tabular}{|c|c|c|c|c|c|}
\hline
\multicolumn{2}{|c|}{\multirow{3}{*}{}}                                                                                         & \multicolumn{2}{c|}{\textbf{Player 2 : Classifier}}                                                                                  & \multirow{2}{*}{\textbf{Threshold}}           & \multirow{2}{*}{\textbf{\begin{tabular}[c]{@{}c@{}}No. of \\ Game \\ Steps\end{tabular}}} \\ \cline{3-4}
\multicolumn{2}{|c|}{}                                                                                                          & \textit{Original}                                                    & \textit{Retrain}                                              &                                               &                                                                                           \\ \cline{3-6} 
\multicolumn{2}{|c|}{}                                                                                                          & \textit{M\_\{acc\}}                                                  & \textit{M\_\{acc\}}                                           & if \textless{}= 50 \%                         &                                                                                           \\ \hline
\multirow{5}{*}{\textbf{\begin{tabular}[c]{@{}c@{}}Player 1:\\ Adversarial \\ Attack\end{tabular}}} & \textit{FGSM}             & 33\%                                                                 & \textbf{68.3\%}                                               & \multirow{2}{*}{}                             & \multirow{2}{*}{\textbf{1}}                                                               \\ \cline{2-4}
                                                                                                    & \textit{One Pixel Attack} & 56\%                                                                 & \textbf{71\%}                                                 &                                               &                                                                                           \\ \cline{2-6} 
                                                                                                    & \multicolumn{1}{l|}{}     & \textit{\begin{tabular}[c]{@{}c@{}}Value from\\ Step 1\end{tabular}} & \textit{\begin{tabular}[c]{@{}c@{}}Retrain \\ 2\end{tabular}} & \multicolumn{1}{l|}{if \textgreater{}= 85 \%} & \multicolumn{1}{l|}{}                                                                     \\ \cline{2-6} 
                                                                                                    & \textit{FGSM}             & 68.3\%                                                               & \textbf{88.5\%}                                               & \multirow{2}{*}{}                             & \multirow{2}{*}{\textbf{2}}                                                               \\ \cline{2-4}
                                                                                                    & \textit{One Pixel Attack} & 71\%                                                                 & \textbf{91.1\%}                                               &                                               &                                                                                           \\ \hline
\end{tabular}
\end{adjustbox}
\label{table1}
\end{table}

We consider that the Adversary is only allowed to perturb the positive instances in the first attack. To calculate the \textit{Utility}, we used Accuracy,$M_{acc}$ as matrice to solve the game. The accuracy matrice is the prediction values by the Model($M$), which is usually determined in percentages, but here we ranged it from $[0-10]\epsilon \mathbb{R}$ for easy computation.

For instance, from Table~\ref{table1}, in the first step after the \textit{FGSM} attack, the classifiers $M_{acc} = 33 \%$. So the gain for the Classifier as $Player_2 (D)$  is $3.3$, which is converted in the range $[0-10]$ in the Kuhn tree Fig~\ref{fig4}. Similarly, for the $Player_1 (A)$, we deduct the value from $100$ as per probability rule, thus the attacker's \textit{Utility} value  $67\%$ in Table~\ref{table1} and $6.6$ in the Kuhn tree.

Now to calculate the Optimal value, we will follow the bottom-up approach. Generally, we will observe the Attacker's perspective, and based on that, the proper Defence will be taken. Fig~\ref{fig5} illustrates the representation of the computation for the optimal value. As it is a two Stage game, after getting the optimal values from the terminal nodes of two binary sub-tree, we get an \textit{Incomplete information Subgame}. The Equilibrium solution for the game is Nash Equilibrium in Table~\ref{table2}. We get unique Nash equilibrium in pure strategies in $(1.15,8.85)$. Therefore, from the Attacker's perspective, it will obtain the optimal value if it chooses FGSM as its attack strategy. Conceding this game-theoretical approach, the Classifier will benefit by estimating the Attacker's possible attack strategies and can take prior precautions.

\begin{table}[!h]
\centering
\caption{Pure Strategy Nash Equilibrium for the first stage Incomplete information game}
%\begin{adjustbox}{width=\textwidth}
\begin{adjustbox}{width=.4\linewidth}
\begin{tabular}{c|c|c}
\hline
\multicolumn{1}{l|}{}                                                                                     & \multicolumn{2}{c}{\textit{\textbf{Player 2 (Classifier)}}} \\ \hline
\multirow{2}{*}{\textit{\textbf{\begin{tabular}[c]{@{}c@{}}Player 1\\ (Adversarial Attack)\end{tabular}}}} & 6.7, 3.3                & \textbf{1.15, 8.85}                \\ \cline{2-3} 
                                                                                                           & 5.6, 4.4                & 0.89, 9.11                         \\ \hline
\end{tabular}
\end{adjustbox}
\label{table2}
\end{table}

In real life, DL models are highly vulnerable to other attacks like PGD \cite{madry2017towards}, Carlini-Wagner attack \cite{carlini2017towards}, FGM attack\cite{dong2018boosting}, etc., and continuously different types of attacks are formulated by attackers. Therefore, it is impossible to track all the attacks and take defensive measures based on that. So understanding the Attacker's rational behavior is essential.

\section{Conclusion}
%\vspace{-3mm}\section{Conclusion}\vspace{-1mm}
In this paper, we introduced a thoroughly analysis of the vulnerability of the CNN model in the Adversarial Attack for the CAPTCHA dataset while emphasizing on possible defense mechanisms. We evaluate the model from a Game theoretical perspective for the optimal solution. Our model is sustainable in the implemented scenario, but in reality, there are other types of Attacks that are not considered in here. So our next target is to extend the game into more than two attacker's perspectives. And also, we need to perform defence strategy like GAN \cite{goodfellow2014explaining}in same dataset to evaluate the defender's performance in multiple attacks.

\bibliographystyle{ieeetr}
\bibliography{reference}

\iffalse

\fi

\end{document}